\documentclass[amssymb,prd,superscriptaddress,aps,nofootinbib,twocolumn,showpacs]{revtex4-1}
\usepackage{graphicx, epsfig, amssymb} %include figure files
\usepackage{amsmath, amsfonts}
\usepackage{bm} %include bold math: \bm{} creates bold letters in math mode
\usepackage[breaklinks]{hyperref}
\usepackage{enumerate}

\usepackage{graphicx,amssymb,amsmath,amsthm,amsfonts,epsfig,epsf}
\usepackage[usenames]{color}
\usepackage{bm}
\usepackage{dcolumn}
\usepackage[latin1]{inputenc}
\usepackage{latexsym}
\usepackage{rotating}
\usepackage{hyperref}
\usepackage{color}
\usepackage{longtable}
\usepackage{enumerate}
\usepackage{tensor}
\usepackage{mathtools}
\usepackage{url}
\def\nn{\nonumber}

\newcommand{\ben}{\begin{enumerate}}
\newcommand{\een}{\end{enumerate}}

\def\be{\begin{equation}}
\def\ee{\end{equation}}
\def\bea{\begin{eqnarray}}
\def\eea{\end{eqnarray}}
\newcommand{\beq}{\begin{eqnarray}}
\newcommand{\eeq}{\end{eqnarray}} 
\newcommand{\ba}{\begin{align}}
\newcommand{\ea}{\end{align}}

\def\nn{\nonumber}

\begin{document}

\title{Linear stability of nonbidiagonal black holes in massive gravity}

\author{Eugeny Babichev}\email{eugeny.babichev@th.u-psud.fr }
%\affiliation{Laboratoire de Physique Th\'eorique d'Orsay, B\^atiment 210, Universit\'e Paris-Sud 11, F-91405 Orsay Cedex, France}
\affiliation{Laboratoire de Physique Th\'eorique, CNRS, Univ. Paris-Sud, Universit\'e Paris-Saclay, 91405 Orsay, France}
%%%%
\author{Richard Brito}\email{richard.brito@tecnico.ulisboa.pt}
\affiliation{CENTRA, Departamento de F\'{\i}sica, Instituto Superior T\'ecnico, Universidade de Lisboa, Avenida~Rovisco Pais 1, 1049 Lisboa, Portugal.}
%%%%
\author{Paolo Pani}\email{paolo.pani@roma1.infn.it}
\affiliation{Dipartimento di Fisica, ``Sapienza'' Universit\`a di Roma \& Sezione INFN Roma1, Piazzale Aldo Moro 5, 00185, Roma, Italy.}
\affiliation{CENTRA, Departamento de F\'{\i}sica, Instituto Superior T\'ecnico, Universidade de Lisboa, Avenida~Rovisco Pais 1, 1049 Lisboa, Portugal.}
%%%

\begin{abstract}
We consider generic linear perturbations of a nonbidiagonal class of static black-hole solutions in massive (bi)gravity.
We show that the quasinormal spectrum of these solutions coincides with that of a Schwarzschild black hole in general relativity, thus proving that these solutions are mode stable. This is in contrast to the case of bidiagonal black-hole solutions which are affected by a radial instability.
On the other hand, the full set of perturbation equations is generically richer than that of a Schwarzschild black hole in general relativity, and this affects the linear response of the black hole to external perturbations.
Finally, we argue that the generalization of these solutions to the spinning case does not suffer from the superradiant instability, despite the fact that the theory describes a massive graviton.
\end{abstract}

\pacs{
% 04.70.-s,04.80.-y,12.60.-i,11.10.St}
%14.80.-j 	Other particles (including hypothetical)
%11.10.St 	Bound and unstable states; Bethe-Salpeter equations
%12.60.-i 	Models beyond the standard model (for unified field theories, see 12.10.-g)
%04.25.D-    Numerical relativity
%04.25.dc    Numerical studies of critical behavior, singularities, and cosmic censorship
%04.25.dg    Numerical studies of black holes and black-hole binaries
04.25.-g    %general relativity: approximation methods, equations of motion
%04.50.-h    Higher-dimensional gravity and other theories of gravity
%04.50.Cd    Kaluza?lein theories
%04.50.Gh    Higher-dimensional black holes, black strings, and related objects
%04.60.Cf    Gravitational aspects of string theory
04.70.-s    %Physics of black holes
%04.70.Bw    Classical black holes
%04.70.Dy    Quantum aspects of black holes, evaporation, thermodynamics
%04.80.-y    Experimental studies of gravity
04.80.Cc     %Experimental tests of gravitational theories
%11.25.Mj    Compactification and four-dimensional models
%11.10.Kk    Field theories in dimensions other than four
}

\maketitle

%%%%%%%%%%%%%%%%%%%%%%%%%%%%%%%%%%%%%%
\section{Introduction}
%%%%%%%%%%%%%%%%%%%%%%%%%%%%%%%%%%%%%%
In addition to passing experimental tests and possibly solving some long-standing issues of general relativity (GR), alternative theories of gravity (cf. Refs.~\cite{Clifton:2011jh,Gair:2012nm,Yunes:2013dva,Berti:2015itd} for some reviews) need also to pass theoretical tests. The latter include internal theoretical consistency, absence of pathologies, and existence of stable gravitational solutions describing physical systems. In this context, black-hole solutions are the ideal test bed to probe the strong-curvature regime of any relativistic (classical) theory of gravity~\cite{Berti:2015itd}. Thus, viable candidates of modified-gravity theories should possess black-hole solutions and the latter should (presumably) be dynamically stable, at least over the typical observation time scale of astrophysical compact objects. In this paper, we focus on massive gravity~\cite{deRham:2010kj,Hassan:2011hr} (a theory that has reacquired considerable attention in the last years, cf. Refs.~\cite{deRham:2014zqa,Schmidt-May:2015vnx} for two recent reviews) and, in particular, on the stability of black-hole solutions in this theory.

The ghost-free, nonlinear completions of massive gravity describe the interaction of two spin-2 fields, either both dynamical~\cite{Hassan:2011hr} or of which only one is dynamical~\cite{deRham:2010kj}.
Due to the absence of uniqueness theorems and to the presence of two independent metrics, various black-hole solutions exist in massive gravity. The solutions known to date can be classified into two classes (see Ref.~\cite{Babichev:2015xha} for a recent review). 
In the first class the two metrics are proportional to each other, whereas this is not the case in the second class. In the static case, the solutions of the first class can always be written in a (bi-)diagonal form, while for the second class it is not possible to simultaneously diagonalize both metrics. In the following we will refer to these solutions as bidiagonal and nonbidiagonal, respectively.

In Refs.~\cite{Babichev:2013una,Brito:2013wya} it was shown that the bidiagonal Schwarzschild solution is generically unstable against radial perturbations. This instability is equivalent~\cite{Babichev:2013una} to the Gregory-Laflamme instability~\cite{Gregory:1993vy} of a five-dimensional black string. Furthermore, Ref.~\cite{Brito:2013wya} considered generic perturbations of both the Schwarzschild and the slowly-rotating Kerr solution when both metrics are proportional to each other, showing that, besides the radial instability, the Kerr solution is also unstable against a superradiant instability (see Ref.~\cite{Brito:2015oca} for a review on superradiance).
Radial perturbations of nonbidiagonal solutions were considered in Ref.~\cite{Babichev:2014oua} showing that, unlike the bidiagonal case, these solutions are classically stable against radial perturbations. One open problem concerns the modal stability of nonbidiagonal solutions to nonradial perturbations.
In this paper we close this gap by considering generic gravitational perturbations of these solutions. Our main result is the proof that the quasinormal-mode (QNM) spectrum of these solutions is the same as that of a Schwarzschild black hole in GR and, therefore, these solutions are classically mode stable\footnote{By ``modes'' we mean the quasinormal spectrum of perturbations, unlike the more generic perturbations which are also considered in this paper.  Similarly, by modal stability we mean that the quasinormal spectrum of perturbations does not contain unstable modes. Strictly speaking, the modal stability does not necessarily imply the full stability of a solution.} precisely as the Schwarzschild metric. 
Along the way we discuss various peculiar properties of the gravitational perturbations of these solutions.

%%%%%%%%%%%%%%%%%%%%%%%%%%%%%%%%%
\section{Setup}
%%%%%%%%%%%%%%%%%%%%%%%%%%%%%%%%%
The Lagrangian of nonlinear massive (bi-)gravity can be written as follows~\cite{deRham:2010ik,Hassan:2011zd}
\be
{\cal L}=\sqrt{|g|}\left[m_g^2R_g+m_f^2\sqrt{{f}/{g}}\, R_f-2m_v^4\, V\left(g,f\right)\right]\,. \label{biaction}
\ee
Here $R_g$ and $R_f$ are the Ricci scalars corresponding to the two metrics $g_{\mu\nu}$ and $f_{\mu\nu}$, respectively; $m_g^{-2}={16\pi G}$, $m_f^{-2}={16\pi \mathcal{G}}$ are the corresponding gravitational couplings. The quantities $f,g$ denote the determinants of the corresponding metric. The potential can be written as
\be
\label{potential}
V:=\sum_{n=0}^4\,\beta_n V_n\left(\gamma\right)\,, \quad \gamma^{\mu}\,_{\nu}:=\left(\sqrt{g^{-1}f}\right)^{\mu}\,_{\nu}\,,
\ee
where $\beta_n$ are real parameters,
%%%%%
\beq
V_0&=&1\,,\quad
V_1=[\gamma]\,,\quad
V_2=\frac{1}{2}\left([\gamma]^2-[\gamma^2]\right)\,,\nn\\
V_3&=&\frac{1}{6}\left([\gamma]^3-3[\gamma][\gamma^2]+2[\gamma^3]\right)\,,\quad
V_4=\det(\gamma)\,,
\eeq
and the square brackets denote the matrix trace.

The Lagrangian~\eqref{biaction} gives rise to two sets of modified Einstein equations for $g_{\mu\nu}$ and $f_{\mu\nu}$, 
\beq
G_{\mu\nu} +\frac{m_v^4}{m_g^2}T_{\mu\nu}(\gamma)&=&0\,, \label{field_eqs1} \\
\mathcal{G}_{\mu\nu} +\frac{m_v^4}{m_f^2}\mathcal{T}_{\mu\nu}(\gamma)&=&0\,, \label{field_eqs2}
\eeq
where $G_{\mu\nu}$ and $\mathcal{G}_{\mu\nu}$ are the corresponding Einstein tensors for the two  metrics $g_{\mu\nu}$ and $f_{\mu\nu}$, and
\beq
T_{\mu\nu}&=&\sum_{n=0}^3(-1)^n\beta_n g_{\mu\lambda}Y^{\lambda}_{\nu}(\gamma)\,,\\
\mathcal{T}_{\mu\nu}&=&\sum_{n=0}^3(-1)^n\beta_{4-n} f_{\mu\lambda}Y^{\lambda}_{\nu}(\gamma^{-1})\,,
\eeq
with $Y(\gamma)=\sum_{r=0}^n (-1)^r\gamma^{n-r}V_r(\gamma)$~\cite{Hassan:2011zd}.
The Bianchi identity implies the conservation conditions
\be
\nabla_g^{\mu}T_{\mu\nu}(\gamma)=0\,,\quad \nabla_f^{\mu}\mathcal{T}_{\mu\nu}(\gamma)=0\,, \label{bianchi1}\\
\ee
where $\nabla_g$ and $\nabla_f$ are the covariant derivatives with respect to $g_{\mu\nu}$ and $f_{\mu\nu}$, respectively. 

%%%%%%%%%%%%%%%%%%%%%%%%%%%%%%%%%
\subsection{Nonbidiagonal spherically symmetric solutions}
%%%%%%%%%%%%%%%%%%%%%%%%%%%%%%%%%

Two classes of static black-hole solutions in this theory can be conveniently written in the bi-advanced Eddington-Finkelstein form~\cite{Babichev:2014oua}
\begin{eqnarray}\label{sol}
ds_g^2 & = & -\left(1-\frac{r_g}{r}\right)dv^2 +2dvdr+r^2 d\Omega^2,\label{metricg}\\
ds_f^2 & = & C^2\left[-\left(1-\frac{r_f}r \right)dv^2 +2dvdr+r^2 d\Omega^2\right], \label{metricf}
\end{eqnarray}
where $C$ is a constant conformal factor and $r_g$ and $r_f$ are the two (generically different) horizon radii of the two metrics.
The only nondiagonal terms of $T^{\mu}_{\phantom{\mu}\nu}$ and $\mathcal{T}^\mu_{\phantom{\mu}\nu}$ read
\begin{equation}\label{Toff}
T^r_{\phantom{r}v} =-C^4\mathcal{T}^r_{\phantom{r}v}=\frac{C \left(\beta_1+2C\beta_2+C^2\beta_3\right) \left(r_g-r_f\right)}{2 r}\,.
\end{equation}
%%%
Clearly, these off-diagonal terms must vanish for the metrics~\eqref{metricg} and~\eqref{metricf} to be solutions of the vacuum field equations.
This implies either $r_g=r_f$, which is equivalent to the (bidiagonal) bi-Schwarzschild solution analyzed in~\cite{Babichev:2013una,Brito:2013wya}, or 
\begin{equation}
\label{relation} 
	\beta_1+2C\beta_2+C^2\beta_3 = 0.
\end{equation} 
%%%
The above condition fixes the value of the conformal factor $C$ for a given choice of the coupling constants $\beta_i$. In the rest of the paper we will focus on this case\footnote{For a full perturbation analysis of the bidiagonal case see Refs.~\cite{Babichev:2013una,Brito:2013wya}.}, which describes two metrics that cannot be simultaneously diagonalized. 
The case of a flat (Minkowski) fiducial metric $f_{\mu\nu}$ coupled to a Schwarzschild metric $g_{\mu\nu}$ falls within this class of solutions.  
The solution~(\ref{sol}) is not the most general analytic nonbidiagonal solution. As it has been shown in Ref.~\cite{Volkov:2012wp}  there is a family of 
nonbidiagonal solutions which contains a function satisfying a nonlinear partial differential equation (see also Ref.~\cite{Volkov:2014ooa}). 
Each regular solution of the partial differential equation gives a different solution for the metrics. 
We consider asymptotically-flat solutions, which implies a fine tuning of the coupling constants such that the two effective cosmological constants vanish. This imposes
%%%%
\beq
\beta_0 &=&-\left(3C\beta_1+3 C^2\beta_2+C^3\beta_3\right)\,,\\
\beta_4 &=&-\frac{\beta_1+3C\beta_2+3C^2\beta_3}{C^3}\,,
\eeq
%%%
to balance the corresponding contributions coming from 
$T_{\mu\nu}$ and $\mathcal{T}_{\mu\nu} $. 

%%%%%%%%%%%%%%%%%%%%%%%%%%%%%%%%%
\section{Gravitational perturbations of the nonbidiagonal solution}
%%%%%%%%%%%%%%%%%%%%%%%%%%%%%%%%%

Let us now consider linear perturbations around the solutions \eqref{metricg} and \eqref{metricf} with the condition~\eqref{relation}, i.e. we focus on nonbidiagonal solutions.
We consider perturbations of the form:
\be
g_{\mu\nu}=g^{(0)}_{\mu\nu}+h_{\mu\nu}^{(g)}\,,\qquad
f_{\mu\nu}=f^{(0)}_{\mu\nu}+h_{\mu\nu}^{(f)}\,,
\ee
where the superscript $^{(0)}$ denotes background quantities and $h_{\mu\nu}$ are small perturbations of the background solutions. The tensors $h_{\mu\nu}^{(g)}$ and
$h_{\mu\nu}^{(f)}$ satisfy the linearized equations 
\begin{equation}\label{perteqs}
	\delta G_{\mu\nu}+  
	\frac{m_v^4}{m_g^2}\delta T_{\mu\nu}=0\,,\qquad 
	\delta \mathcal{G}_{\mu\nu} + \frac{m_v^4}{m_f^2}\delta\mathcal{T}_{\mu\nu}=0\,,
\end{equation} 
%%%
where $G_{\mu\nu}(g)=G_{\mu\nu}^{(0)}+\delta G_{\mu\nu}$, $\mathcal{G}_{\mu\nu}(f)=\mathcal{G}_{\mu\nu}^{(0)}+\delta \mathcal{G}_{\mu\nu}$ and similarly for $\delta T_{\mu\nu}$ and $\delta\mathcal{T}_{\mu\nu}$.

In a spherically symmetric background, the spin-2 perturbations $h_{\mu\nu}^{(g)}$ and
$h_{\mu\nu}^{(f)}$ can be decomposed in terms of axial and polar perturbations. In Fourier space, this decomposition schematically reads:
\be
\label{decom}
h_{\mu\nu}(v,r,\theta,\phi)=\frac{1}{\sqrt{2\pi}}\sum_{l,m}\int_{-\infty}^{+\infty}d\omega e^{-i\omega v} \tilde h_{\mu\nu}^{lm}(\omega,r,\theta,\phi)\,,
\ee
for both $h_{\mu\nu}^{(g)}$ and $h_{\mu\nu}^{(f)}$, where $\tilde h_{\mu\nu}^{lm}:=h^{{\rm axial},lm}_{\mu\nu}+h^{{\rm polar},lm}_{\mu\nu}$, $h^{{\rm axial},lm}_{\mu\nu}$ and $h^{{\rm polar},lm}_{\mu\nu}$ are explicitly given in Appendix~\ref{app:decomposition}.  Without loss of generality, we will also multiply the definition of  $h_{\mu\nu}^{(f)}$ by an overall $C^2$ factor. Spherical symmetry assures that the field equations do not depend on the azimuthal number $m$. In addition, perturbations with different parity and different harmonic index $l$ decouple from each other\footnote{To simplify the notation, we shall often omit the superscript ${}^{lm}$ in the perturbation functions.}. 

In the nonbidiagonal case~\eqref{relation}, by using this decomposition, it turns out that the mass terms in the perturbation equations~\eqref{perteqs} take the remarkably simple form
\begin{widetext}
\begin{align}\label{nbd}
&\delta T^{\mu}_{\phantom{\mu}\nu}=
\frac{\mathcal{A} \left(r_g-r_f\right) }{4 r}\, e^{-i \omega v}\nonumber\\
&\begin{pmatrix}
 0 & 0 & 0 & 0 \\
 2K^{lm}_{(-)} Y_{lm} & 0 & -\left(h^{lm}_{1(-)}\frac{\partial_{\phi}Y_{lm}}{\sin\theta}+\eta^{lm}_{1(-)}\partial_{\theta}Y_{lm}\right) & h^{lm}_{1(-)}\sin\theta\partial_{\theta}Y_{lm}-\eta^{lm}_{1(-)}\partial_{\phi}Y_{lm} \\
 -\left(h^{lm}_{1(-)}\frac{\partial_{\phi}Y_{lm}}{\sin\theta}+\eta^{lm}_{1(-)}\partial_{\theta}Y_{lm}\right) & 0 & H^{lm}_{2(-)} Y_{lm} & 0 \\
 \frac{1}{r^2\sin\theta}\left(h^{lm}_{1(-)}\partial_{\theta}Y_{lm}-\eta^{lm}_{1(-)}\frac{\partial_{\phi}Y_{lm}}{\sin\theta}\right) & 0 & 0 & H^{lm}_{2(-)} Y_{lm}
\end{pmatrix}\,,
\end{align}
\end{widetext}
and $C^4\delta\mathcal{T}^{\mu}_{\phantom{\mu}\nu}=-\delta T^\mu_{\phantom{\mu}\nu}$, where $h_{\mu\nu}^{(-)} := h_{\mu\nu}^{(f)}- h_{\mu\nu}^{(g)}$ and $\mathcal{A} = 2C^2\left(\beta_2+C\beta_3\right)$.
%%%%%

By taking the divergence of Eq.~\eqref{perteqs} and using the Bianchi identities for the Einstein tensors, 
we obtain the constraint 
$ \nabla^\nu_{(f)}\delta \mathcal{T}_{\mu\nu}\propto \nabla^\nu_{(g)}\delta T_{\mu\nu} =0$, which, from Eq.~\eqref{nbd} in the nonbidiagonal case ($r_g\neq r_f$), yields the following relations:
\beq\label{divg}
&&H^{lm}_{2(-)}=0\,,\\
&&\left(r \eta^{lm}_{1(-)}\right)'=0\,,\\ 
&&\left(r K^{lm}_{(-)}\right)'+\frac{l(l+1)\eta^{lm}_{1(-)}}{2r}=0\,,\\
&&\left(r h^{lm}_{1(-)}\right)'=0\,.
\eeq
The above equations can be immediately solved for
\beq
&&H^{lm}_{2(-)}=0\,,\qquad \eta^{lm}_{1(-)} = \frac{c_0}{r}\,, \label{constnbd1} \\
&& h^{lm}_{1(-)} = \frac{c_1}{r}\,,\qquad K^{lm}_{(-)}= \frac{c_2}{r}+\frac{l(l+1)c_0}{2r^2} \label{constnbd2} \,,
\eeq
where $c_0$, $c_1$, and $c_2$ are (generically complex) integration constants\footnote{We should note that, since we are working in the frequency-domain, $c_i$ are arbitrary functions of $\omega$ while in the time-domain they are arbitrary (real) functions of the advanced time $v$.}. The peculiar structure of $\delta T^{\mu}_{\phantom{\mu}\nu}$ is responsible for some highly nontrivial properties which are discussed in the section below.

%%%%%%%%%%%%%%%%%%%%%%%%%%%%%%%%%
\section{Equivalence of the QNMs to those of a Schwarzschild black hole in GR}\label{sec:QNMs}
%%%%%%%%%%%%%%%%%%%%%%%%%%%%%%%%%
In this section we show that the QNMs of the nonbidiagonal black-hole solution of massive gravity are the same as those of a Schwarzschild black hole in GR.

The QNMs are the proper frequencies of vibration of a relativistic self-gravitating object, in analogy with the normal modes of oscillating stars in Newtonian gravity (cf. Refs.~\cite{Kokkotas:1999bd,Berti:2009kk,Konoplya:2011qq} for some reviews). Due to the emission of gravitational waves or to absorption by an event horizon, the QNMs are complex numbers whose real part defines the frequency of the perturbation, whereas the imaginary part defines the inverse of the damping time (or of the instability time scale in the case of unstable modes). It should be stressed that the QNMs do not form a complete set~\cite{Kokkotas:1999bd} so they do not describe the full response of the black hole to external perturbations. In fact, we show below that QNMs of nonbidiagonal black holes in bigravity are exactly the same as in GR, while generic perturbations of bigravity black holes are different from those of GR black holes.

The QNMs can be computed as the eigenvalues of a boundary-value problem defined by Eq.~\eqref{perteqs} with suitable boundary conditions. For the case of static, asymptotically-flat black holes, regularity imposes that the perturbations behave as ingoing waves near the horizon, $\sim e^{-i (\omega t+k_- r_*)}$ and as outgoing waves near infinity, $\sim e^{i (k_+ r_*-\omega t)}$. Here, $r_*$ is the tortoise coordinate defined through $v=t+r_*$, where $t$ is a Schwarzschild-like time coordinate\footnote{For clarity, in this Section the metric perturbations $h_{\mu\nu}$ are written as functions of $t$. One can always do this by defining $v=t+r_*$ and, after substituting in Eq.~\eqref{decom}, absorb the $e^{-i\omega r_*}$ factor into the inverse-Fourier transformed quantities $\tilde h_{\mu\nu}$.}. The constant $k_\pm$ (which we assume to be positive without loss of generality) is the momentum of the perturbations and it is related to the effective dispersion relation. For example, for an outgoing perturbation with effective mass $\mu$ propagating in Minkowski spacetime\footnote{For gravitational perturbations of GR Schwarzschild black holes $k_\pm=\pm\omega$, whereas for the static bidiagonal black-hole solutions of massive gravity $k_-=\omega$ and $k_+=\sqrt{\omega^2-\mu^2}$, consistently with the propagation of a massive mode.}, $k_+=\sqrt{\omega^2-\mu^2}$.

Therefore, the QNMs of the bimetric system are defined by the following boundary conditions for the metrics $h^{(g)}_{\mu\nu}$ and $h^{(f)}_{\mu\nu}$,
%%%
\begin{eqnarray}
   \tilde{h}^{(g)}_{\mu\nu} \to A_{\mu\nu}^\pm e^{\pm i k_\pm r_{g*}}\,, \quad   \tilde{h}^{(f)}_{\mu\nu} \to B_{\mu\nu}^\pm e^{\pm i k_\pm r_{f*}} \,,\label{BCs}
\end{eqnarray}
%%%
where $A_{\mu\nu}^\pm$ and $B_{\mu\nu}^\pm$ are typically polynomials in $1/r$, the plus (minus) sign refers to the near-infinity (near-horizon) behavior, whereas the
tortoise coordinates are defined via $dr/dr_{g*}=\left(1-r_g/r\right)$ and $dr/dr_{f*}=\left(1-r_f/r\right)$.

Inspection of Eqs.~\eqref{constnbd1} and \eqref{constnbd2} together with the decomposition~\eqref{decom} immediately shows that the boundary conditions~\eqref{BCs} cannot be satisfied unless $c_i=0$ in Eqs.~\eqref{constnbd1} and \eqref{constnbd2}. For example, from Eqs.~\eqref{constnbd1}, \eqref{constnbd2}, \eqref{decom},~\eqref{oddpart} and~\eqref{evenpart}, we obtain
\begin{equation}
 \tilde h^{(f)}_{r\phi}-\tilde  h^{(g)}_{r\phi} = e^{-i\omega r_*} \left[\frac{c_0}{r}\partial_\phi Y_{lm}-\frac{c_1}{r}\sin\theta\partial_{\theta}Y_{lm}\right]\,,
\end{equation}
for the difference of the inverse-Fourier transformed quantities $\tilde h^{(f)}_{r\phi}$ and $\tilde h^{(g)}_{r\phi}$ (and similarly for other components). Therefore, it is clear that the difference $\tilde h^{(f)}_{r\phi}-\tilde  h^{(g)}_{r\phi}$ represents an ingoing wave of frequency $\omega$ in the whole space and the same property must hold independently for $\tilde h^{(f)}_{r\phi}$ and $\tilde h^{(g)}_{r\phi}$. Because $r_{g*}\sim r_{f*}\to-\infty$ near the corresponding horizon, the near-horizon boundary condition in Eq.~\eqref{BCs} is always satisfied with $k_-=\omega$. On the other hand, the near-infinity boundary condition,  $h^{(g)}_{\mu\nu}\sim h^{(f)}_{\mu\nu}\to e^{ik_+r}$, cannot be enforced\footnote{If we were using retarded Eddington-Finkelstein coordinates, the opposite situation would occur: the solution would describe an outgoing wave in the whole space, and the boundary conditions would be automatically satisfied at infinity but not at the event horizon. In both cases, the full set of boundary conditions~\eqref{BCs} cannot be enforced unless $c_i=0$.}. 

This simple observation implies that the boundary conditions for QNMs impose $c_0=c_1=c_2=0$ and, in turn, $\delta T^\mu_\nu=\delta {\cal T}^\mu_\nu=0$\footnote{In the special case ${\cal A}=0$, i.e., $\beta_2=-C \beta_3$, one always gets $\delta T^\mu_\nu=\delta {\cal T}^\mu_\nu=0$ and the perturbation equations reduce to the standard linearized Einstein's equations as noted in Ref.~\cite{Kobayashi:2015yda} (see also~\cite{Kodama:2013rea}  for the case with only one dynamical metric). This can be also related to an extra symmetry for spherically symmetric solutions in the case $\beta_2=-C \beta_3$~\cite{Volkov:2012wp,Babichev:2014fka}.}. Therefore, the eigenvalue problem reduces to the standard linearized Einstein's equations
%%%
\begin{equation}
	\delta G_{\mu\nu}=0\ , \ \ \ \delta \mathcal{G}_{\mu\nu} =0\ ,\label{perteqs0}
\end{equation} 
%%%
with the extra constraints coming from Eqs.~\eqref{constnbd1} and \eqref{constnbd2} with $c_0=c_1=c_2=0$ , namely
%%%
\beq
H^{lm}_{2(g)}=H^{lm}_{2(f)}\,,&\quad & \eta^{lm}_{1(g)} = \eta^{lm}_{1(f)} \,,\label{constnbd1b} \\
h^{lm}_{1(g)} = h^{lm}_{1(f)}\,,&\quad & K^{lm}_{(g)}=K^{lm}_{(f)} \,.\label{constnbd2b} 
\eeq
%%%

To complete our proof, we can use the freedom to choose a particular gauge. In this case it is convenient to choose a gauge such that $H^{lm}_{2(g)}=K^{lm}_{(g)}=\eta^{lm}_{1(g)}=h^{lm}_{1(g)}=0$. This can always be imposed by transforming~\cite{Zerilli:1971wd}
\begin{equation}
 h_{\mu\nu}^{(g)} \to h_{\mu\nu}^{(g)} -\nabla_{\mu}\xi_{\nu}-\nabla_{\nu}\xi_{\mu}\,,
\end{equation}
where  $\xi_\mu$ is the transformation four-vector. The latter can be decomposed into an axial vector component and into three polar vector components, which can be chosen to enforce the aforementioned relations $h^{lm}_{1(g)}=0$ and $H^{lm}_{2(g)}=K^{lm}_{(g)}=\eta^{lm}_{1(g)}=0$, respectively.
Since there is only one diffeomorphism invariance and two metrics, the components of the metric $f$ are not fixed  {\it a priori} by the above gauge choice.
%%%
However, Eqs.~\eqref{constnbd1b} and~\eqref{constnbd2b} imply $H^{lm}_{2(f)}=K^{lm}_{(f)}=\eta^{lm}_{1(f)}=h^{lm}_{1(f)}=0$. Therefore, Eq.~\eqref{perteqs0} reduces to two copies of the linearized Einstein equations in the gauge $H^{lm}_{2}=K^{lm}=\eta^{lm}_{1}=h^{lm}_{1}=0$. Note that this gauge is different from the standard Regge-Wheeler-Zerilli gauge, in which $G^{lm}_{2}=\eta^{lm}_{0}=\eta^{lm}_{1}=h^{lm}_{2}=0$~\cite{Regge:1957td,Zerilli:1971wd}. Nonetheless, the perturbation equations are precisely the same as in the case of GR.

Thus, we have just proved that the eigenvalue problem reduces to that of two Schwarzschild metrics with horizon radii $r_g$ and $r_f$ in GR. In particular, there will be no monopole and dipole modes, the QNMs exist only for $l\geq2$, and they correspond to 2 propagating degrees of freedom. 
%%%
As a by-product of this equivalence, the QNM spectrum does not contain any unstable mode and the nonbidiagonal black-hole solution of massive gravity is therefore mode stable for any gravitational perturbations (which can be decomposed into quasinormal modes). 
Both properties (the absence of $l=0$ and $l=1$ modes and the modal stability) are in striking contrast to the case of bidiagonal solutions~\cite{Brito:2013wya,Brito:2015oca}, as we also discuss in the next section.

%%%%%%%%%%%%%%%%%%%%%%%%%%%%%%%%%
\section{Generic Gravitational Perturbations}
%%%%%%%%%%%%%%%%%%%%%%%%%%%%%%%%%
As shown in the previous section, the QNM spectrum of the nonbidiagonal black hole in massive gravity coincides with that of a Schwarzschild black hole in GR. This property is true for both the axial and polar sectors, which respectively reduce to a Regge-Wheeler and a Zerilli equation. Nonetheless, the full set of perturbations (and therefore the object's response to external sources) is generically different, both in the axial and in the polar sector. In this section we relax the boundary conditions at infinity to include ingoing (at infinity) perturbations, unlike the previous section where those perturbations were forbidden by boundary conditions corresponding to QNMs. Thus, our study will include more general perturbations which are useful to study the linear response of the black hole to external perturbers. 
In the following we will consider the axial and polar sectors, and the cases $l=0$, $l=1$ and $l\geq2$, separately. 

%%%%%%%%%%%%%%%%%%%%%%%%%%%%%%%%%
\subsection{Polar sector}
%%%%%%%%%%%%%%%%%%%%%%%%%%%%%%%%%
Here we discuss the perturbation equations for the polar sector separately for $l=0$, $l=1$ and $l\geq2$.

%%%%%%%%%%%%%%%%%%%%%%%%%%%%%%%%%
\subsubsection{Polar monopole}\label{ssec:monopole}
%%%%%%%%%%%%%%%%%%%%%%%%%%%%%%%%%

Radial (i.e., $l=0$) perturbations were studied in Ref.~\cite{Babichev:2014oua}. In this case the perturbation functions $G^{lm}$, $\eta^{lm}_{0}$ and $\eta^{lm}_{1}$ are not defined because their corresponding angular part in Eq.~\eqref{evenpart} vanishes. For $\omega=0$, one gets $c_0=0$ and $h_{\mu\nu}^{(f)}=h_{\mu\nu}^{(g)}$ and there is one solution which corresponds to a trivial mass shift in both metrics $f_{\mu\nu}$ and $g_{\mu\nu}$. When $\omega\neq 0$, we find the same solution as in Ref.~\cite{Babichev:2014oua} when using the same gauge. For the sake of completeness, we here show the explicit form of this solution.

Unlike Ref.~\cite{Babichev:2014oua}, however, let us choose a gauge such that $H_{1(g)}=K_{(g)}=H_{2(g)}=0$. 
%%%%
From Eqs.~\eqref{constnbd1} and \eqref{constnbd2} we then have $H_{2(f)}=0$ and $K_{(f)}=c_2/r$. Finally, the field equations yield 
\begin{align}
&H_{1(f)}=i c_5\omega\,,\\
&H_{0(f)}=-i c_2\omega-\frac{c_2 r_f}{2 r^2}-2i\omega c_5\left(1-\frac{r_f}{r}\right)\nonumber\\
&+\frac{\mathcal{A}c_2 m_v^4(r_f-r_g)}{4m_f^2 C^2i\omega r}\,,\\
&H_{0(g)}=\frac{\mathcal{A}c_2 m_v^4(r_g-r_f)}{4m_g^2 i\omega r}\,,
\end{align}
where $c_5$ is an integration constant\footnote{Note that the result in \cite{Babichev:2014oua} is written in terms of $h^{\mu\nu}$, while here we work with $h_{\mu\nu}$, hence the apparent difference of the expressions.}. 
%%%
There are two free integration constants, $c_2$ and $c_5$, which are not fixed by the assumption of asymptotic flatness. This can be checked by 
calculating the curvature invariants. For example, the Kretschmann scalar $R_{abcd}R^{abcd}$ of the $l=0$ polar solution vanishes at large distances for any value of the integration constants. Moreover, in this gauge $c_5$ does not affect the $g_{\mu\nu}$ metric, and does not contribute either to the curvature of both metrics or to the energy-momentum tensors $\delta T_{\mu\nu}$ and $\delta\mathcal{T}_{\mu\nu}$. 

In other words, if one takes the $g_{\mu\nu}$ metric to be the physical one and couples it to matter, the constant $c_5$ would be completely decoupled and would not affect any observable physical quantity, at least to linear order (we discuss possible nonlinear effects in Sec.~\ref{sec:conclusion}).

On the contrary, the constant $c_2$ is physical. This constant cannot be gauged away from either of the metrics, contributes to $\delta T_{\mu\nu}$ and $\delta\mathcal{T}_{\mu\nu}$, and is therefore associated with observable quantities.

For any $c_2\neq 0$, due to the term $e^{-i\omega v}$ appearing in Eq.~\eqref{decom}, the solution above describes an ingoing wave which does not feel any effective potential and therefore does not change its propagation in the entire space.
This property is reminiscent of Minkowski spacetime, in which an ingoing wave is not backscattered due to the absence of a gravitational potential\footnote{The analogy with the Minkowski spacetime extends also to the computation of QNMs previously discussed. Minkowski spacetime does not possess proper modes of vibration due to the absence of an effective potential. However, one could imagine to add a test, perfectly-absorbing surface at some fixed location $r=r_0$, which would play the role of an event horizon. Similarly to what previously discussed, in this case one can impose purely absorbing boundary conditions at $r=r_0$ but it would be impossible to impose simultaneously the correct boundary conditions at infinity. Due to the absence of backscattering, Minkowski spacetime does not possess  QNMs even in the presence of a perfectly absorbing surface.}. 
This behavior is in contrast to the Schwarzschild case in GR, in which the radial mode is nondynamical. On the other hand, the radial perturbations of the bidiagonal metric are described by a Zerilli-like equation~\cite{Brito:2013wya}
%%%
\begin{equation}
  \frac{d^2 \Psi}{dr_{g*}^2}-2i\omega \frac{d\Psi}{dr_{g*}}- V_0(r)\Psi=0\,,
\end{equation}
%%%%
where the effective potential $V_0(r)$ is given below Eq.~(30) in Ref.~\cite{Brito:2013wya}. As discussed in Refs.~\cite{Babichev:2013una,Brito:2013wya}, not only in this case is the perturbation dynamical, but it also leads to an instability.

%%%%%%%%%%%%%%%%%%%%%%%%%%%%%%%%%
\subsubsection{Polar dipole}
%%%%%%%%%%%%%%%%%%%%%%%%%%%%%%%%%
\label{ssec:dipole}
When $l=1$, the function $G^{lm}$ is not defined. Up to gauge freedom we can set $H_{2(g)}=\eta_{1(g)}=K_{(g)}=0$. By using the constraints~\eqref{constnbd1} and \eqref{constnbd2}, a straightforward calculation then leads to the following solution

\begin{eqnarray}
H_{0(f)}&=& 2 r c_9-i \omega  c_2\nn\\
&&-\frac{\left(2 c_9+\omega  \left(\omega  c_2+(2 i-4 r \omega ) c_6+6 i r c_9\right)\right) r_f}{2 r^2 \omega ^2}\nn\\
&&-\frac{i {\cal A} m_v^4 \left(\omega  c_0+(3 r\omega-i ) c_2\right) }{12 C^2 r^2 \omega ^2 m_f^2}\left(r_f-r_g\right)\,,\\
H_{0(g)}&=& \frac{\omega  (2 r \omega -i) c_7 r_g+c_8 \left(2 r^3 \omega
   ^2-(1+3 i r \omega ) r_g\right)}{r^2 \omega ^2} \nn\\
&&+\frac{{\cal A} m_v^4 \left(c_2+i \omega  \left(c_0+3 r c_2\right)\right) }{12 r^2 \omega ^2 m_g^2}\left(r_f-r_g\right) \,,\\
H_{1(f)}&=& c_6\,,\\
H_{1(g)}&=& c_7\,,\\
\eta_{0(f)}&=& \frac{i \omega  c_0}{2} +\frac{c_2}{2}+r \left(c_6+r c_9\right)+\frac{\left(i \omega  c_6+c_9\right) r_f}{r \omega ^2}\nn\\
&&+\frac{{\cal A} m_v^4 \left(i \omega  c_0+c_2\right)}{12 C^2 r \omega ^2 m_f^2}\left(r_f-r_g\right)\,,\\
\eta_{0(g)}&=& r \left(c_7+r c_8\right)+\frac{\left(i \omega  c_7+c_8\right) r_g}{r \omega ^2}\nn\\
&&-\frac{{\cal A} m_v^4 \left(i \omega  c_0+c_2\right) }{12 r \omega ^2 m_g^2}\left(r_f-r_g\right)\,,
\end{eqnarray}
where $c_i$ are integration constants. 
The perturbations must be small in order to stay within the validity of the perturbation theory, i.e. $h^{(g)}_{\mu\nu}\ll g_{\mu\nu}$, and similar for perturbations of the second metric. This requirement leads to $c_6=c_7=c_8=c_9=0$. The only free constants are then $c_0$ and  $c_2$. Both these constants induce physical (observable) changes in the metric perturbations, unlike the monopole case, where only one constant is physical and the other one is a gauge constant. 
Nevertheless, similarly to the monopole case, this solution describes a purely ingoing wave which is not backscattered by the black hole.

Also in this case the GR solution describes a gauge mode and is nondynamical, whereas the $l=1$ polar sector of the bidiagonal solution describes two propagating degrees of freedom governed by a pair of coupled equations (cf. Eqs.(44) and (45) in Ref.~\cite{Brito:2013wya}). Contrary to the nonbidiagonal case under consideration, the bidiagonal solution possesses $l=1$ polar QNMs which depend on the graviton mass~\cite{Brito:2013wya}.

%%%%%%%%%%%%%%%%%%%%%%%%%%%%%%%%%
\subsubsection{Polar perturbations with $l\geq 2$}
%%%%%%%%%%%%%%%%%%%%%%%%%%%%%%%%%
\label{ssec:quadropole}
The $l\geq2$ polar case is qualitatively similar to the $l\geq2$ axial case (considered below) although technically more involved. Also in this case we can adopt a gauge such that  $H^{lm}_{2(g)}=K^{lm}_{(g)}=\eta^{lm}_{1(g)}=0$ which, from Eqs.~\eqref{constnbd1} and \eqref{constnbd2}, implies  $H^{lm}_{2(f)}=0$, $K^{lm}_{(f)}=c_2/r+l(l+1)c_0/(2r^2)$ and $\eta^{lm}_{1(f)}=c_0/r$. After some algebra, the field equations can be solved for $H_{1(g)}^{lm}$, $H_{1(f)}^{lm}$, $G_{(g)}^{lm}$, $G_{(f)}^{lm}$, ${d\eta_{0(g)}^{lm}}/dr$ and ${d\eta_{0(f)}^{lm}}/dr$, whereas the functions $H_{0(g)}^{lm}$ and $H_{0(f)}^{lm}$ satisfy a set of two decoupled, second-order differential equations, namely
%%%%
%
\beq
{\cal D}_g [{\tilde \Phi}_g]-W_g {\tilde \Phi}_g &=& s_g\,,\label{polarg}\\
{\cal D}_f [{\tilde \Phi}_f]-W_f {\tilde \Phi}_f &=& s_f\,, \label{polarf}
\eeq
where ${\tilde \Phi}_g:=r^2 H_{0(g)}^{lm} /(r-r_g)$ and ${\tilde \Phi}_f:=r^2 H_{0(f)}^{lm} /(r-r_f)$ and we defined the differential operators
%%%
\begin{equation}
 {\cal D}_{g}=\frac{d^2}{dr_{g*}^2}-2i\omega \frac{d}{dr_{g*}} \,,\quad {\cal D}_{f}=\frac{d^2}{dr_{f*}^2}-2i\omega \frac{d}{dr_{f*}}\,.\\
\end{equation}
%%%%
In the above equations, the potentials read
\beq
W_g&=&\frac{l (l+1) \left(r-r_g\right)-2 i r \omega  \left(2 r-3 r_g\right)+r_g}{r^3}\,,\\
W_f&=&\frac{l (l+1) \left(r-r_f\right)-2 i r \omega  \left(2 r-3 r_f\right)+r_f}{r^3}\,,
\eeq
whereas the source terms are
\begin{widetext}
 \beq
s_g&=& \frac{{\cal A} m_v^4 r \left[c_0 \Lambda+2 c_2 r\right] \left(r_g-r_f\right)-4 B_1 m_g^2 \left(r_g+2 i r^2
   \omega \right)+4 c_4 r m_g^2 \left[r_g \left(\Lambda+4 i r \omega \right)+2 i \left(\Lambda-2\right) r^2
   \omega \right]}{4 r^3 m_g^2}\,,\\
   %%%%
s_f&=& \frac{{\cal A} m_v^4 \left(c_0 \Lambda+2 c_2 r\right) \left(r_f-r_g\right)}{4 C^2 r^2 m_f^2}\nn\\
&+&\frac{2 r^2 \omega 
   \left(-2 i B_2+2 i c_3 \left(\Lambda-2\right) r+c_0 \Lambda \omega +2 c_2 r \omega \right)-r_f \left(2
   B_2-2 c_3 r \left(\Lambda+4 i r \omega \right)+i c_0 \Lambda \omega +c_2 (2+6 i r \omega )\right)}{2
   r^3}\,,
\eeq
\end{widetext}
%%%%
where $\Lambda:=l(l+1)$ and $B_1$ and $B_2$ are two further integration constants. Similar to the previous cases, the validity of the perturbation theory requires $c_2=c_3=c_4=0$, otherwise the functions $G_{(g)}^{lm}$, $G_{(f)}^{lm}$, $\eta_{0(g)}^{lm}$ and $\eta_{0(f)}^{lm}$ would grow linearly with $r$ at large distances. 
Note that Eqs.~\eqref{polarg} and \eqref{polarf} are decoupled from each other and, in the GR limit\footnote{The source terms vanish when ${\cal A}=0$ and when the integration constants $c_i$ and $B_i$ are set to zero. In the GR case, this choice can be done without loss of generality.}, they reduce to two copies of the same homogeneous differential equation. The latter is not in the standard Zerilli form~\cite{Zerilli:1970se} but, quite interestingly, is precisely the Bardeen-Press-Teukolsky equation for gravitational perturbations of the Schwarzschild metric in GR~\cite{Bardeen:1973xb,Teukolsky:1972my,Teukolsky:1973ha} (cf. Eq.~(5.2) in Ref.~\cite{Teukolsky:1973ha} when the black-hole spin is zero). It is easy to check that this equation is isospectral to the Regge-Wheeler equation by performing a Chandrasekhar transformation~\cite{Chandra:1975xx} (see Appendix~\ref{app:Teu}). We have also checked this property numerically by transforming the homogeneous equations into a 4-term recurrence relation and by computing the modes through continued fractions~\cite{Berti:2009kk,Pani:2013pma}.

As for the $l\geq2$ axial case that we discuss below, the source terms $s_g$ and $s_f$ in Eqs.~\eqref{polarg} and \eqref{polarf} do not alter the QNM spectrum, in agreement with the generic argument presented in Sec.~\ref{sec:QNMs}. The situation is therefore rather different from that of the bidiagonal solution~\cite{Brito:2013wya}. In the latter case, the $l\geq2$ polar perturbations reduce to a set of three coupled ordinary-differential equations (cf. Eqs.~(38)--(40) in Ref.~\cite{Brito:2013wya}, which propagate three degrees of freedom and correspond to a quasinormal spectrum that depends on the graviton mass.
%%%%

%%%%%%%%%%%%%%%%%%%%%%%%%%%%%%%%%
\subsection{Axial sector}
%%%%%%%%%%%%%%%%%%%%%%%%%%%%%%%%%
The axial sector does not contain a monopole ($l=0$) and one is left with the axial dipole mode ($l=1$) and with the higher multipoles $l\geq2$, which we treat separately.

%%%%%%%%%%%%%%%%%%%%%%%%%%%%%%%%%
\subsubsection{Axial dipole mode}
%%%%%%%%%%%%%%%%%%%%%%%%%%%%%%%%%

When $l=1$, the angular functions $W_{lm}$ and $X_{lm}$ in Eq.~\eqref{oddpart} vanish (and therefore $h^{lm}_{2}$ is not defined), while $h_{1(f)}=c_1/r+h_{1(g)}$ from Eq.~\eqref{constnbd2}. The $(v,\theta)$ component of the field equations~\eqref{perteqs} yields
\beq
r^2 h''_{0(g)}&=&2 h_{0(g)}-i r \omega \left(r h'_{1(g)}+2 h_{1(g)}\right)\,, \label{eqs_axialdi1}\\
r^2 h''_{0(f)}&=&2 h_{0(f)}-i r \omega \left(r h'_{1(g)}+2 h_{1(g)}+c_1/r\right)\,.\label{eqs_axialdi2}
\eeq
The residual gauge freedom can be used to set one of the axial functions to zero. If we impose $h_{1(g)}=0$, from the constraints~\eqref{constnbd1} and \eqref{constnbd2} we obtain $h_{1(f)}=c_1/r$ and in such case Eqs.~\eqref{eqs_axialdi1} and \eqref{eqs_axialdi2} can be solved for
\begin{align}
&h_{0(g)}=r^2 c_3+\frac{i(r_g-r_f) c_1\mathcal{A}m_v^4}{12 m_g^2 \omega  r}\,,\\
&h_{0(f)}=r^2 c_4+\frac{i(r_f-r_g) c_1\mathcal{A}m_v^4}{12 m_f^2C^2\omega r} +\frac{i c_1\omega}{2}\,.\label{h0f}
\end{align}
where $c_3$ and $c_4$ are two further integration constants. 
The constants of integration $c_3$ and $c_4$ must be set to zero, otherwise the perturbative approach breaks down at large $r$.
On the other hand, $c_1\neq 0$ does not violate our (relaxed) assumptions on the metrics: indeed both metrics are asymptotically flat, as can be checked by computing curvature invariants. In particular, the Pontryagin density ${}^{*}\!RR:=\frac{1}{2}\epsilon^{abef}R_{abcd}R^{cd}{}_{ef}\sim c_1/r^7$. 
Note that $f_{\mu\nu}$ is asymptotically flat but not Minkowski in this case, due to the last term in Eq.~\eqref{h0f}. 

This solution is qualitatively similar to the $l=0$ polar case.
As in the $l=0$ polar case, the above solution represents a purely ingoing wave which does not feel any effective potential and, therefore, it is not backscattered by the geometry. 

Thus, the axial dipolar perturbation of the nonbidiagonal black-hole solution describes a dynamical (purely ingoing) wave. As such, this solution cannot be an eigenfunction of the boundary-value problem, in agreement with our previous analysis which showed that no dipolar QNMs exist for this solution. Nonetheless, this behavior is dramatically different from the case of GR --~in which the $l=1$ mode is pure gauge and therefore nondynamical~-- and also from the case of the bidiagonal solution. In the latter case, the dipolar axial sector is described by a single second-order Regge-Wheeler-like equation~\cite{Brito:2013wya},
%%%
\begin{equation}
 \frac{d^2 \tilde\Psi}{dr_{g*}^2}-2i\omega \frac{d\tilde\Psi}{dr_{g*}}-\left(1-\frac{r_g}{r}\right)\left(\mu^2+\frac{6}{r^2}-\frac{8 r_g}{r^3}\right)\tilde\Psi=0\,,
\end{equation}
%%%%
where $\mu$ is the effective mass of the propagating mode (which is proportional to the graviton mass in the theory) and $\Psi(v,r)\sim e^{-i\omega v} \tilde\Psi(\omega,r)$ is a master function. It is easy to show that, in this case, the general solution describes a superposition of outgoing and ingoing waves both at the horizon and at infinity and that the eigenvalue problem admits a novel set of QNMs~\cite{Brito:2013wya}.

%%%%%%%%%%%%%%%%%%%%%%%%%%%%%%%%%
\subsubsection{Axial perturbations with $l\geq 2$}
%%%%%%%%%%%%%%%%%%%%%%%%%%%%%%%%%

In this case, to simplify the equations, we define two new radial functions given by 
\beq
{\tilde Q}_g=r^3\left(\frac{h^{lm}_{0(g)}}{r^2}\right)'+i\omega r h^{lm}_{1(g)}\,, \label{Qg}\\
{\tilde Q}_f=r^3\left(\frac{h^{lm}_{0(f)}}{r^2}\right)'+i\omega r h^{lm}_{1(f)}\,.
\eeq
From the $(v,\theta)$ component of the field equations~\eqref{perteqs}, we can then obtain two algebraic equations for $h^{lm}_{2(g)}$ and $h^{lm}_{2(f)}$, which allow us to eliminate these functions from the other equations. From the $(r,\theta)$ components, we get two second-order differential equations for ${\tilde Q}_g$ and ${\tilde Q}_f$, namely 
\beq
{\cal D}_g [{\tilde Q}_g]-V_g {\tilde Q}_g &=&\frac{c_1 m_v^4 \mathcal{A}}{2 m_g^2 r^2}\left(r_g-r_f\right)\left(1-\frac{r_g}{r}\right),\label{axialg}\\
{\cal D}_f [{\tilde Q}_f]-V_f {\tilde Q}_f &=&\frac{c_1 m_v^4 \mathcal{A}}{2 m_f^2  r^2 C^2}\left(r_f-r_g\right)\left(1-\frac{r_f}{r}\right), \label{axialf}
\eeq
where the effective potentials read
\beq
V_g&=&\left(1-\frac{r_g}{r}\right)\left[\frac{l(l+1)}{r^2}-\frac{3r_g}{r^3}\right]\,, \label{Vg}\\
V_f&=&\left(1-\frac{r_f}{r}\right)\left[\frac{l(l+1)}{r^2}-\frac{3r_f}{r^3}\right]\,. \label{Vf}
\eeq
Note that the field equations allow us to compute only the master functions $\tilde Q_g$ and $\tilde Q_f$ and not the metric perturbations $h_{0(g)}$, $h_{1(g)}$ and $h_{0(f)}$, $h_{1(f)}$  separately. This is consistent with the existence of a residual gauge freedom. For example, the function $h_{1(g)}$ in Eq.~\eqref{Qg} can be set to zero through a gauge choice. In such a case, $h_{1(f)}=c_1/r$ from Eq.~\eqref{constnbd2}.

Note that, when $c_1=0$, Eqs.~\eqref{axialg} and \eqref{axialf} reduce to a pair of Regge-Wheeler equations~\cite{Regge:1957td}, and are thus identical to the case of GR, consistent with our general argument in Sec.~\ref{sec:QNMs}.
On the other hand, the terms proportional to $c_1$ act as a source of the Regge-Wheeler equation and cannot modify the proper frequencies of the system. This is discussed in more detail in Sec.~\ref{sec:time} below.

Also in this case it is interesting to compare the perturbations of the nonbidiagonal solutions with those of a Schwarzschild black hole in GR and with those of the bidiagonal solution of massive gravity. In the former case, the perturbation describes a single propagating degree of freedom governed by Eq.~\eqref{axialg} with $c_1=0$. In the latter case, the $l\geq2$ axial sector is described by two propagating degrees of freedom, but they are governed by a \emph{coupled} system of equations (cf. Eqs.~(32) and~(33) in Ref.~\cite{Brito:2013wya}) which are also associated with a different set of quasinormal frequencies. Finally, the perturbation equations in the bidiagonal case depend on the graviton mass, similar to the $l=1$ case previously discussed, whereas the graviton mass in the nonbidiagonal case appears only in the source terms, but not in the effective potentials~\eqref{Vg} and \eqref{Vf} (the same property holds true in the $l\geq2$ polar case discussed above).

%%%%%%%%%%%%%%%%%%%%%%%%%%%%%%%%%
\subsection{Time evolution} \label{sec:time}
%%%%%%%%%%%%%%%%%%%%%%%%%%%%%%%%%
In this section we consider the time evolution governed by the perturbation equation~\eqref{axialg} [or, equivalently, Eq.~\eqref{axialf}] in the time domain, in order to investigate the role of the source term appearing on the right-hand side of this equation. A similar analysis for the $l\geq2$ polar sector is more technically involved but it is qualitatively similar. (As shown in Appendix~\ref{app:Teu}, for the polar case the only difference is in the source term. One can show that although the sources are more complicated, their asymptotic behavior at the horizon and at infinity is similar to the axial case, and thus the waveforms are qualitatively similar.)

By introducing a new radial function $\tilde{Z}_g=e^{-i\omega r_{*g}}{\tilde Q}_g$, Eq.~\eqref{axialg} becomes
\be
\frac{d^2 \tilde{Z}_g}{dr_{g*}^2}+\left(\omega^2-V_g\right)  \tilde{Z}_g=\left(1-\frac{r_g}{r}\right)S_g\,,\label{axialZg0}\\
\ee
where 
\be\label{source_evolution}
S_g=e^{-i\omega r_{g*}}c_1\left(r_g-r_f\right)\frac{m_v^4 \mathcal{A}}{2 m_g^2 r^2}\,.
\ee

As previously discussed, the above source term appears in the perturbed nonbidiagonal solution and it would vanish in the case of GR. To investigate the impact of such a term on the waveform, we assume that the latter is produced by a driving force at $t=0$, which for simplicity we take to be a static Gaussian. In the frequency-domain this amounts to adding a source to the right-hand side of Eq.~\eqref{axialZg0}, namely
\be\label{source_evolution_2}
S_{\rm Gaussian}=A_0 e^{-\left(r_{*g}-r_0\right)^2/2\sigma^2}\,.
\ee

Thus, the full time-evolution equation reads
\be
\frac{d^2 \tilde{Z}_g}{dr_{g*}^2}+\left(\omega^2-V_g\right)  \tilde{Z}_g=\left(1-\frac{r_g}{r}\right)S\,,\label{axialZg}\\
\ee
where $S:=S_g+(1-r_g/r)^{-1}S_{\rm Gaussian}$.
To obtain the waveform $Z_g(t,r)$\footnote{Note that due to Eq.~\eqref{decom} and the definition $\tilde{Z}_g=e^{-i\omega r_{*g}}{\tilde Q}_g$, $Z_g(t,r)$ is a function of $t:= v-r_{*g}$.} we use an inverse-Fourier transform,
\be\label{waveform_td}
Z_g(t,r)=\frac{1}{\sqrt{2\pi}}\int_{-\infty}^{+\infty} e^{-i\omega t}\tilde{Z}_g(\omega,r)d\omega\,,
\ee
where $\tilde{Z}_g(\omega,r)$ is computed using the Green's function technique outlined in Appendix~\ref{app:GF}. In principle, $c_1$ is an arbitrary function of $\omega$ which depends on the initial conditions of the perturbations $h_1$. For simplicity, here we consider the case where $c_1$ is a constant, which is sufficient for our argument. For this choice, in the time domain, the source~\eqref{source_evolution} is proportional to the Dirac delta function $\delta(v)$.

\begin{figure}[th]
\begin{center}
%\begin{tabular}{c}
\epsfig{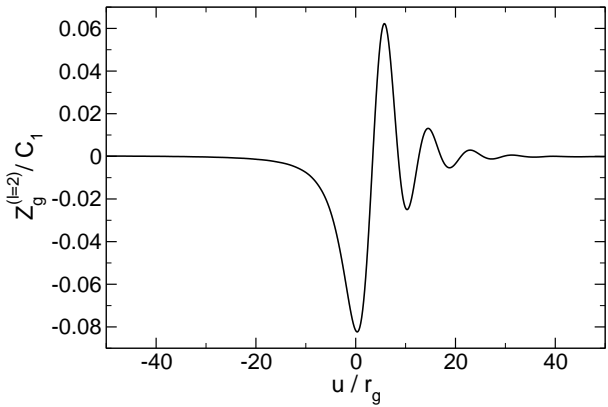}
%\end{tabular}
\end{center}
\caption{\label{fig:waveform} The waveform $Z_g(t,r)$ (in units of $C_1:=c_1\left(r_g-r_f\right)\frac{m_v^4 \mathcal{A}}{2 m_g^2}$) with $l=2$ as a function of $u:= t-r_{*g}$ (in units of $r_g$). This is the solution of Eq.~\eqref{axialZg0}, i.e. in the case where the external perturbation $S_{\rm Gaussian}$ is absent. It is easy to check that the ringdown signal is governed by the QNMs of a Schwarzschild black hole in GR.
}
\end{figure}

Let us first consider the case in which no external source is present, i.e. we solve Eq.~\eqref{axialZg} with $A_0=0$ [or, equivalently, Eq.~\eqref{axialZg0}]. In this case the waveform is proportional to the combination $C_1:=\frac{c_1 m_v^4 \mathcal{A} }{2 m_g^2}\left(r_g-r_f\right)$. The waveform obtained with the Green's function method is shown in Fig.~\ref{fig:waveform}\footnote{We work in units where $G=c=1$. In these units, the constant $C_1:= c_1\left(r_g-r_f\right)m_v^4 \mathcal{A}/(2 m_g^2)$ is dimensionless.}. A straightforward Fourier analysis of the waveform shows that the ringdown signal~\cite{Berti:2009kk} is governed precisely by the QNMs of a Schwarzschild black hole in GR. This is natural since Eq.~\eqref{axialZg0} is equivalent to the standard Regge-Wheeler equation in GR but with an external source term given by $S_g$. As in the case of a forced harmonic oscillator, the source can modify the waveform but not the proper modes of the system (for a similar analysis in a different modified theory of gravity, see Ref.~\cite{Molina:2010fb}), which are still described by the QNMs of the solution, i.e. by the same QNMs of a Schwarzschild black hole in GR.

\begin{figure}[th]
\begin{center}
%\begin{tabular}{c}
\epsfig{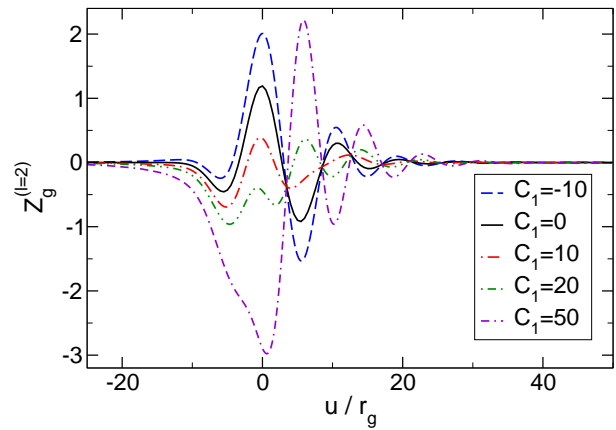}
%\end{tabular}
\end{center}
\caption{\label{fig:waveform_Gaussian} Waveform $Z_g(t,r)$ for $l=2$ as a function of $u:= t-r_{*g}$ and for different values of $C_1$. This is a solution of the full time-evolution equation~\eqref{axialZg} with an external Gaussian source~\eqref{source_evolution_2}, with  $A_0 r_g^2=0.4$, $r_0=5r_g$ and $\sigma=2.5 r_g$. From this waveform it is easy to check that, for any value of $C_1$, the ringdown signal is governed by the QNMs of a Schwarzschild black hole in GR. }
\end{figure}

As is clear from the above discussion, adding an external source term like Eq.~\eqref{source_evolution_2} is simply equivalent to solving the standard Regge-Wheeler equation in GR but with an effective source term given by $S$ in Eq.~\eqref{axialZg}. The waveform obtained by solving Eq.~\eqref{axialZg} for different values of $C_1$ and for a representative external source term is shown in Fig.~\ref{fig:waveform_Gaussian}. Also in this case a straightforward frequency decomposition shows that, for any value of $C_1$, the ringdown waveform is governed by the QNMs of the Schwarzschild solution in GR, although the black-hole response to the external perturbation depends on $C_1$. This is in agreement with our proof given in Sec.~\ref{sec:QNMs}.

% 

%%%%%%%%%%%%%%%%%%%%%%%%%%%%%%%%%
\subsection{Perturbations of slowly-rotating Kerr}
%%%%%%%%%%%%%%%%%%%%%%%%%%%%%%%%%

Our results show that, unlike in the bidiagonal case, gravitational perturbations of nonbidiagonal static black holes in massive gravity do not allow for (quasi)-bound states. The latter are long-lived modes trapped in the potential well that typically develops for massive perturbations (cf. e.g. Refs.~\cite{Dolan:2007mj,Brito:2013wya,Brito:2015oca}). This is due to the fact that: (i) the perturbations with $l=0,1$ do not feel any effective potential and (ii) perturbations with $l\geq2$ propagate exactly in the effective potential of a Schwarzschild black hole in GR; in particular, such effective potential does not depend on the graviton mass. 

One of the consequences of bosonic quasi-bound states in the spectrum is the existence of a superradiant instability~\cite{Brito:2015oca} that affects the spinning bidiagonal black-hole solutions in massive gravity~\cite{Brito:2013wya}. Indeed, a stable long-lived mode can turn unstable in the spinning case due to Zeeman splitting of the quasinormal frequencies~\cite{Pani:2013pma,Brito:2015oca}.  

A generalization of the nonbidiagonal solution~\eqref{sol} describing a rotating black hole was found in Ref.~\cite{Babichev:2014tfa} (a further generalization describing the Kerr-(anti-)de Sitter black holes was presented in Ref.~\cite{Ayon-Beato:2015qtt}). Due to the absence of quasi-bound states in the static case, for this family of solutions our results strongly suggest that no superradiant instability exists, at least in the slowly-rotating regime.
%

%%%%%%%%%%%%%%%%%%%%%%%%%%%%%%%%%
\section{Conclusion and discussion}\label{sec:conclusion}
%%%%%%%%%%%%%%%%%%%%%%%%%%%%%%%%%
We derived the full set of linearized equations governing gravitational perturbations of the nonbidiagonal Schwarzschild solution in massive (bi)gravity. We showed that the quasinormal spectrum of these solutions coincides with that of a Schwarzschild black hole in GR. This result is quite surprising and has some interesting consequences. In general, massive (bi)gravity propagates more degrees of freedom than GR (including massive modes), so one might naively expect that black-hole solutions possess more modes of vibration and that the latter would depend on the value of the graviton mass. This is indeed the case for bidiagonal solutions~\cite{Babichev:2013una,Brito:2013wya}, but it is not the case for the nonbidiagonal solutions discussed here. 

Furthermore, the bidiagonal solution possesses an unstable radial mode, which is absent in the nonbidiagonal case~\cite{Babichev:2014oua}\footnote{Due to the instability of the bidiagonal solutions along with the existence of several other spherically symmetric solutions~\cite{Volkov:2012wp,Brito:2013xaa,Babichev:2015xha}, the outcome of gravitational collapse in massive gravity is still unclear (see also Ref.~\cite{Enander:2015kda} for arguments showing that gravitational collapse of stars might not lead to black-hole formation in these theories).}. Finally, massive bosonic perturbations generically allow for quasi-bound, long-lived modes in the spectrum of spherically-symmetric black holes. Such modes can turn (superradiantly) unstable when the black hole rotates above a certain threshold~\cite{Brito:2015oca}. Remarkably, such long-lived modes are absent in the static nonbidiagonal solution. This suggests that, when spinning, this solution does not suffer from the superradiant instability. If this conjecture is confirmed, the nonbidiagonal black-hole solution of massive gravity would be the first case of a spinning black-hole geometry which is mode-stable in a theory that propagates a massive bosonic field.

It is also natural to conjecture that the QNMs of the rotating black hole found in Ref.~\cite{Babichev:2014tfa} are identical to those of a Kerr black hole in GR, similar to the static case discussed in this paper.

The absence of extra QNMs, compared to GR, naturally raises the question about the number of propagating degrees of freedom 
on top of nonbidiagonal black holes. In particular, one may worry about the disappearance of some degrees of freedom and, consequently, possible strong coupling.
Indeed, some modes are indeed absent, as compared to, e.g. bidiagonal black holes. 
At the same time, there appear modes, which do not feel any potential and therefore do not backscatter. 
The absence of backscattering implies that these modes do not satisfy the boundary conditions imposed for QNMs.
Nevertheless these ``free propagating'' modes depend on the initial conditions and their impact on the resulting metric cannot be removed by a gauge transformation.
We would like to stress here that these perturbations contain free functions, as ``normal'' propagating modes do, 
and initial conditions are required to impose them.
Indeed, each integration constant, e.g. $c_0$ and $c_2$  in Sec.~\ref{ssec:dipole}  are functions of $\omega$ and 
when converted to the time domain they yield free functions. 
Note that, on the contrary, in the special case $\beta_2=-C \beta_3$ studied in Refs.~\cite{Kodama:2013rea,Kobayashi:2015yda} the ``free propagating'' modes are absent, 
so the solution is certainly strongly coupled in this specific case.
In the general case, however, a separate study is required to find explicitly whether all the modes are truly dynamical, and hence to address the issue of possible strong coupling. 

It would also be interesting to go beyond the linear level, employed in this paper, and to consider nonlinear effects. 
This question is connected to the possible strong coupling issue. If some of the degrees of freedom happen not to propagate on the background of nonbidiagonal black-hole solutions (due to their peculiarity) one would naturally expect that at least some of them reappear at the nonlinear level. 
If this is indeed the case, then the nonbidiagonal solutions may be nonlinearly unstable. 
Nonlinear effects may change our discussion in Section~\ref{ssec:monopole}, where we argued that one of the two integration constants is a pure gauge, since it can be reabsorbed in the perturbations $f_{\mu\nu}$ and it does not give a contribution to the mass energy-momentum tensor. This constant might source physical perturbations of $g_{\mu\nu}$ at the nonlinear level, thus activating a physical degree of freedom. Nonlinear effects may also generate a potential for those modes which propagate from infinity down to the horizon without scattering. 

We should also mention that our study did not address the question of ghosts in the spectrum of perturbations 
since it relies on the analysis of the field equations. This issue may be addressed together with the question about the number of propagating degrees of freedom mentioned above, for example, by a Hamiltonian analysis.

%%%%%%%%%%%%%%%%%%%%%%%%%%%%%%%%%%%%%%%%%%%%%%%%%%%%%%%%%%%%%%%%%%%%
\begin{acknowledgments}
We thank Vitor Cardoso, Alessandro Fabbri and Mikhail Volkov for interesting comments.
E.B. was supported by the research program ``Programme national de cosmologie et galaxies of the CNRS/INSU'', France and Russian Foundation for Basic Research Grant No. RFBR 15-02-05038.
R.B. acknowledges financial support from the FCT-IDPASC program through the grant SFRH/BD/52047/2012.
P.P. was supported by the European Community through
the Intra-European Marie Curie Contract No.~AstroGRAphy-2013-623439 and by FCT-Portugal through the project IF/00293/2013. 
\end{acknowledgments}
%%%%%%%%%%%%%%%%%%%%%%%%%%%%%%%%%%%%%%%%%%%%%%%%%%%%%%%%%%%%%%%%%%%%

%%%%%%%%%%%%%%%%%%%%%%%%%%%%%%%%%%%%%%%%%%%%%%%%%%%%%%%%%%%%%%%%%%%%%%%%%%%%%%%%%%%%%%%%%%%%%%%%%%%%%%%%%
\appendix
%%%%%%%%%%%%%%%%%%%%%%%%%%%%%%%%%%%%%%%%%%%%%%%%%%%%%%%%%%%%%%%%%%%%%%
\section{Tensor spherical harmonic decomposition of spin-2 fields}\label{app:decomposition}
%%%%%%%%%%%%%%%%%%%%%%%%%%%%%%%%%%%%%%%%%%%%%%%%%%%%%%%%%%%%%%%%%%%%%%

In a spherically symmetric background spin-2 field perturbations can be decomposed in terms of axial and polar quantities and expanded in a complete basis of tensor spherical harmonics. For the expansion~\eqref{decom}, the axial and polar parts are given, respectively, by~\cite{Regge:1957td}
\begin{widetext}
\begin{equation}\label{oddpart}
h^{{\rm axial},lm}_{\mu\nu}(\omega,r,\theta,\phi) =
 \begin{pmatrix}
  0 & 0 & h^{lm}_0(\omega,r)\csc\theta\partial_{\phi}Y_{lm}(\theta,\phi) & -h^{lm}_0(\omega,r)\sin\theta\partial_{\theta}Y_{lm}(\theta,\phi) \\
  * & 0 & h^{lm}_1(\omega,r)\csc\theta\partial_{\phi}Y_{lm}(\theta,\phi) & -h^{lm}_1(\omega,r)\sin\theta\partial_{\theta}Y_{lm}(\theta,\phi) \\
  *  & *  & -h^{lm}_2(\omega,r)\frac{X_{lm}(\theta,\phi)}{\sin\theta} & h^{lm}_2(\omega,r)\sin\theta W_{lm}(\theta,\phi)  \\
  * & * & * & h^{lm}_2(\omega,r)\sin\theta X_{lm}(\theta,\phi)
 \end{pmatrix}\,,
\end{equation}
\begin{align}\label{evenpart}
h^{{\rm polar},lm}_{\mu\nu}(\omega,r,\theta,\phi)=
\begin{pmatrix}
H_0^{lm}(\omega,r)Y_{lm} & H_1^{lm}(\omega,r)Y_{lm} & \eta^{lm}_0(\omega,r)\partial_{\theta}Y_{lm}& \eta^{lm}_0(\omega,r)\partial_{\phi}Y_{lm}\\
  * & H_2^{lm}(\omega,r)Y_{lm} & \eta^{lm}_1(\omega,r)\partial_{\theta}Y_{lm}& \eta^{lm}_1(\omega,r)\partial_{\phi}Y_{lm}\\
  *  & *  & \begin{array}{c}r^2\left[K^{lm}(\omega,r)Y_{lm}\right.\\\left. +G^{lm}(\omega,r)W_{lm}\right]\end{array} & r^2  G^{lm}(\omega,r)X_{lm}  \\
  * & * & * & \begin{array}{c}r^2\sin^2\theta\left[K^{lm}(\omega,r)Y_{lm}\right.\\\left.-G^{lm}(\omega,r)W_{lm}\right]\end{array}
\end{pmatrix}\,,
\end{align}
\end{widetext}
where asterisks represent symmetric components, $Y_{lm}:= Y_{lm}(\theta,\phi)$ are the scalar spherical harmonics and

\be
X_{lm}(\theta,\phi)=2\partial_{\phi}\left[\partial_{\theta}Y_{lm}-\cot\theta Y_{lm}\right]\,,
\ee
\be
W_{lm}(\theta,\phi)=\partial^2_{\theta}Y_{lm}-\cot\theta\partial_{\theta}Y_{lm}-\csc^2\theta\partial^2_{\phi}Y_{lm}\,.
\ee
In the case of massive gravity, the decompositions \eqref{oddpart} and \eqref{evenpart} are applied to $h_{\mu\nu}^{(g)}$ and $h_{\mu\nu}^{(f)}$ (in the latter case with an arbitrary $C^2$ factor in front) with different perturbation functions.

%%%%%%%%%%%%%%%%%%%%%%%%%%%%%%%%%%%%%%%%%%%%%%%%%%%%%%%%%%%%%%%%%%%%%%
\section{Bardeen-Press-Teukolsky equation and Chandrasekhar transformation}\label{app:Teu}
%%%%%%%%%%%%%%%%%%%%%%%%%%%%%%%%%%%%%%%%%%%%%%%%%%%%%%%%%%%%%%%%%%%%%%

One can easily prove that the $l\geq2$ polar QNMs are isopectral to the $l\geq2$ axial QNMs by showing that they are governed by the same equations. 
By defining the radial functions ${\tilde Y}_g:=e^{-i\omega r_{*g}} H_{0(g)}^{lm} /(r-r_g)^2$ and ${\tilde Y}_f:=e^{-i\omega r_{*f}} H_{0(f)}^{lm} /(r-r_f)^2$ we find, after some algebra, the following equations:
%%%%
%
\beq
{\cal L}_g [{\tilde Y}_g] &=& e^{-i\omega r_{*g}}T_g\,,\label{polarg_Teu}\\
{\cal L}_f [{\tilde Y}_f] &=& e^{-i\omega r_{*f}}T_f\,, \label{polarf_Teu}
\eeq
where we defined the differential operators
%%%
\beq
&&{\cal L}_{g}=\left(r^2-r_g r\right)\frac{d^2}{dr^2}+6\left(r-\frac{r_g}{2}\right)\frac{d}{dr}\nonumber\\
&&+\frac{r^4\omega^2-4ir^2\omega\left(r-\frac{r_g}{2}\right)}{\left(r^2-r_g r\right)}
+8ir\omega-l(l+1)+6\nonumber\\
&&{\cal L}_{f}=\left(r^2-r_f r\right)\frac{d^2}{dr^2}+6\left(r-\frac{r_f}{2}\right)\frac{d}{dr}\nonumber\\
&&+\frac{r^4\omega^2-4ir^2\omega\left(r-\frac{r_f}{2}\right)}{\left(r^2-r_f r\right)}
+8ir\omega-l(l+1)+6\,,
\eeq
%%%%
whereas the source terms are
\begin{widetext}
 \beq
&&T_g=\frac{c_0{\cal A}\Lambda m_v^4 (r_g-r_f)}{4 m_g^2 r (r-r_g)^2}+\frac{c_2 {\cal A} m_v^4 (r_g-r_f)}{2 m_g^2 (r-r_g)^2}
+c_4\frac{\Lambda r_g+2 i r \omega  \left(\left(\Lambda-2\right) r+2 r_g\right)}{r (r-r_g)^2}-B_1\frac{r_g+2 i r^2 \omega }{r^2 (r-r_g)^2}\,,\\
   %%%%
&&T_f= \frac{c_0\Lambda \left({\cal A} m_v^4 r (r_f-r_g)+2 C^2 m_f^2 \omega  \left(2 r^2 \omega -i r_f\right)\right)}{4 C^2
   m_f^2 r^2 (r-r_f)^2}+c_2\frac{{\cal A}  m_v^4 r^2 (r_f-r_g)+2 C^2 m_f^2 \left(2 r^3 \omega ^2-r_f (1+3 i r \omega )\right)}{2 C^2 m_f^2 r^2
   (r-r_f)^2}\nonumber\\
&&	+c_3\frac{\Lambda r_f+2 i r \omega  \left(\left(\Lambda-2\right) r+2 r_f\right)}{r (r-r_f)^2}
-B_2\frac{r_f+2 i r^2 \omega }{r^2 (r-r_f)^2}\,.
\eeq
\end{widetext}
%%%%
In the GR limit these two equations reduce to two copies of the Bardeen-Press-Teukolsky equation (in the form originally written by them) for gravitational perturbations of the Schwarzschild metric in GR~\cite{Bardeen:1973xb,Teukolsky:1972my,Teukolsky:1973ha}. By performing a Chandrasekhar transformation~\cite{Chandra:1975xx} of the form (as given in Ref.~\cite{Hughes:2000pf}):
\beq
r^2{\tilde Y}_g&=&{\cal D}^2_{-g}\left(r \tilde{X}_g\right)\label{chandra_g}\,,\\
r^2{\tilde Y}_f&=&{\cal D}^2_{-f}\left(r \tilde{X}_f\right)\label{chandra_f}\,,
\eeq
where ${\cal D}_{-g,f}\equiv d/dr-i r \omega/(r-r_{g,f})$, one finds that the functions $\tilde{X}_{g,f}$ satisfy the following Regge-Wheeler equations:
\beq
&&\frac{d^2 \tilde{X}_g}{dr_{g*}^2}+\left(\omega^2-V_g\right)  \tilde{X}_g=S^P_{g}\,,\label{polarXg}\\
&&\frac{d^2 \tilde{X}_f}{dr_{f*}^2}+\left(\omega^2-V_f\right)  \tilde{X}_f=S^P_{f}\,,\label{polarXf}
\eeq
with the effective potentials given in Eqs.~\eqref{Vg} and~\eqref{Vf}, whereas the source terms can be obtained by inserting the transformations~\eqref{chandra_g} and~\eqref{chandra_f} in Eqs.~\eqref{polarg_Teu} and~\eqref{polarf_Teu}, using~\eqref{polarXg} and~\eqref{polarXf} to eliminate $X_{g,f}$ and their derivatives, and then solving for $S^P_{g,f}$, which can be found analytically. Since their analytical expression is rather lengthy and their explicit form is not fundamental we do not show it here. Similar to the axial case, to the leading order the source terms decay as $S^P_{g,f}\sim 1/r^2$ when $r\to\infty$ [cf. Eq.~\eqref{source_evolution}]. 
Comparing this with the axial case, Eq.~\eqref{axialZg0}, one immediately sees that the only difference is in the source term, and thus under the same boundary conditions, the QNM spectrum of the polar and axial sector is the same (and coincides with that of a GR Schwarzschild black hole).

%%%%%%%%%%%%%%%%%%%%%%%%%%%%%%%%%%%%%%%%%%%%%%%%%%%%%%%%%%%%%%%%%%%%%%
\section{Green's function}\label{app:GF}
%%%%%%%%%%%%%%%%%%%%%%%%%%%%%%%%%%%%%%%%%%%%%%%%%%%%%%%%%%%%%%%%%%%%%%

The Green's function $G_{l\omega}$ of Eq.~\eqref{axialZg} is defined by
\be
\frac{d^2 G_{l\omega}}{dr_{g*}^2}+\left(\omega^2-V_g\right)  G_{l\omega}=\delta(r_{g*}-r'_{g*})\,.
\ee
To construct the Green's function we choose two independent solutions of the homogeneous equation associated with Eq.~\eqref{axialZg}, $\tilde{Z}_g^{H}$ and $\tilde{Z}_g^{\infty}$, which satisfy the following boundary conditions:
\begin{equation} \label{boundinf}
\left\{
 \begin{array}{l}
 \tilde{Z}_g^{\infty}\sim e^{i\omega r_{g*}}\,,\\
\tilde{Z}_g^{H}\sim A_{\rm{out}}e^{i\omega r_{g*}}+A_{\rm{in}}e^{-i\omega r_{g*}}\,,  
\end{array}\right.
 \quad r_{g*}\to +\infty
\end{equation}
\begin{equation}\label{boundhor}
\left\{
 \begin{array}{l}
\tilde{Z}_g^{\infty}\sim B_{\rm{out}}e^{i\omega r_{g*}}+B_{\rm{in}} e^{-i\omega r_{g*}}\,,\\
\tilde{Z}_g^{H}\sim e^{-i\omega r_{g*}}\,,
\end{array}\right.
 \quad r_{g*}\to -\infty\,,
\end{equation}
where $\{A,B\}_{\rm{in},\rm{out}}$ are constants. 
By imposing the boundary conditions discussed in Sec.~\ref{sec:QNMs}, the Green's function reads
\be
G_{l\omega}(r'_{g*},r_{g*})=\frac{1}{W}\left\{
 \begin{array}{l}
\tilde{Z}_g^{H}(r_{g*})\tilde{Z}_g^{\infty}(r'_{g*})\,, \quad r_{g*}<r'_{g*}\,,\\
\tilde{Z}_g^{\infty}(r_{g*})\tilde{Z}_g^{H}(r'_{g*})\,, \quad r_{g*}>r'_{g*}\,,
\end{array}\right.
\ee
where $W$ is the Wronskian of these two linearly independent solutions, and it is constant by virtue of the field equation~\eqref{axialZg}. Evaluating $W$ at infinity one gets,
\begin{equation} \label{wronskian}
W=\tilde{Z}_g^{H}\frac{d\tilde{Z}_g^{\infty}}{dr_{g*}}-\tilde{Z}_g^{\infty}\frac{d\tilde{Z}_g^{H}}{dr_{g*}}=2i\omega A_{\rm in}\,.
\end{equation}

The solution to Eq.~\eqref{axialZg} with appropriate boundary conditions is then given by
\be
\tilde{Z}_g(r_{g*})=\int_{-\infty}^{+\infty} dr'_{g*}\,\,G_{l\omega}(r'_{g*},r_{g*})\, S(r'_{g*})\,.
\ee
Evaluating this expression at $r_{g*}\to +\infty$ we find
\beq
&&\tilde{Z}_g(r_{g*}\to\infty)=\frac{\tilde{Z}_g^{\infty}(r_{g*})}{W}\int_{-\infty}^{+\infty} dr'_{g*}\,\,\tilde{Z}_g^{H}(r'_{g*})\, S(r'_{g*})\nn\\
&&=\frac{e^{i\omega r_{g*}}}{2i\omega A_{\rm in}}\int_{r_g}^{+\infty} dr'\,\,\tilde{Z}_g^{H}(r')\, S(r'_{g*})\left(1-\frac{r_g}{r'}\right)^{-1}\,.
\eeq
This integral can be computed numerically by first integrating the homogeneous part of Eq.~\eqref{axialZg} with the boundary condition~\eqref{boundhor} to get $\tilde{Z}_g^{H}$ and then compute $A_{\rm in}$ by equating the solution obtained numerically to~\eqref{boundinf}. The waveform in the time-domain is then obtained performing the integral~\eqref{waveform_td}. For more details on the numerical procedure see, e.g., Ref.~\cite{Cardoso:2002ay}.

%%%%%%%%%%%%%%%%%%%%%%%
\bibliography{ref}  
%%%%%%%%%%%%%%%%%%%%%%%%%%

\end{document}